\documentclass[twocolumn,showpacs,aps,prl,superscriptaddress]{revtex4}

\usepackage{graphicx}
\usepackage{dcolumn}
\usepackage{amsmath}
\usepackage{epsfig}

\input pubboard/babarsym

\newcommand{\BABARPubYear}    {06}
\newcommand{\BABARPubNumber}  {065}

\newcommand{\SLACPubNumber} {12200}

\def\figurebox#1#2#3{%
    \def\arg{#3}%
    \ifx\arg\empty
    {\hfill\vbox{\hsize#2\hrule\hbox to #2{\vrule\hfill\vbox to #1{\hsize#2\vfill}\vrule}\hrule}\hfill}%
    \else
    {\hfill\epsfbox{#3}\hfill}%
    \fi}

\def\Resultphieta{2.1\pm 0.4 (\mathrm{stat})\pm 0.1(\mathrm{syst}) ~\mathrm{fb}}
\def\Resultphietascale{2.9\pm 0.5 (\mathrm{stat})\pm 0.1(\mathrm{syst}) ~\mathrm{fb}}

\begin{document}
\preprint{\babar-PUB-\BABARPubYear/\BABARPubNumber} 
\preprint{SLAC-PUB-\SLACPubNumber}

\begin{flushleft}
\babar-PUB-\BABARPubYear/\BABARPubNumber\\
SLAC-PUB-\SLACPubNumber\\
\end{flushleft}

\title
{
{\large \bf 
Observation of the Exclusive Reaction $e^+e^-\rightarrow$  $\phi\eta$ at $\sqrt{s}=10.58$ GeV
} 	
}

%
\author{B.~Aubert}
\author{M.~Bona}
\author{D.~Boutigny}
\author{Y.~Karyotakis}
\author{J.~P.~Lees}
\author{V.~Poireau}
\author{X.~Prudent}
\author{V.~Tisserand}
\author{A.~Zghiche}
\affiliation{Laboratoire de Physique des Particules, IN2P3/CNRS et Universit\'e de Savoie, F-74941 Annecy-Le-Vieux, France }
\author{E.~Grauges}
\affiliation{Universitat de Barcelona, Facultat de Fisica, Departament ECM, E-08028 Barcelona, Spain }
% ...
\collaboration{The \babar\ Collaboration}
\noaffiliation

\date{\today}

\begin{abstract}

We report the observation of \epem\to$\phi\eta$ near $\sqrt{s}$ = 10.58 GeV
with 6.5 $\sigma$ significance in the $K^+K^-\gamma\gamma$ final state in a data
sample of 224~\invfb  collected by the \babar~ experiment at 
the  \pep2 $e^+e^-$ storage rings.
We measure the restricted radiation-corrected cross section to be 
$\sigma(\epem\!\! \to\! \phi \eta)\! =$$\Resultphieta$ within the range 
$|\cos\theta^*|\! <\! 0.8$, where $\theta^*$ is the center-of-mass polar angle of 
the $\phi$ meson.  The $\phi$ meson is required  
to be in the invariant mass range of
1.008 $<~\!\! m_{\phi}\!\! ~<$ 1.035~\gevcc. The radiation-corrected cross section in the full 
$\cos\theta^*$ range is extrapolated to be $\Resultphietascale$.

\end{abstract}

\maketitle

The large data samples of the B factories provide an opportunity
    to explore rare exclusive quasi-two-body processes in \epem annihilation,
    such as final states produced through one virtual photon with negative
    C-parity ($J/\psi \eta_c$ or  other double charmonium 
states)~\cite{belledoublecc,babardoublecc}, and two-virtual-photon
    annihilation (TVPA) with positive C-parity ($\rho^0\rho^0$ or $\phi\rho^0$)~\cite{babarvv}. 
The process \epem\to$J/\psi \eta_c$ and other double charmonium processes are 
observed at rates approximately ten times larger than the expectation from QCD-based models~\cite{nrqcd}. 
Various theoretical efforts have been made to resolve the discrepancy between experimental and theoretical 
results~\cite{aclightcone,bodwinlightcone,leenewresult}. 
Another avenue to explore this puzzle is provided by the related
    process \epem\to$\phi\eta$.
A recent observation of $\psi(3770) \to \phi\eta$ at a branching fraction 
of $(3.1\pm0.6\pm0.3\pm0.1)\times10^{-4}$~\cite{cleophieta} also stimulates  a search  for $\Upsilon(4S)\to\phi\eta$.

\begin{figure}
\begin{center}
\includegraphics[width=6.5cm]{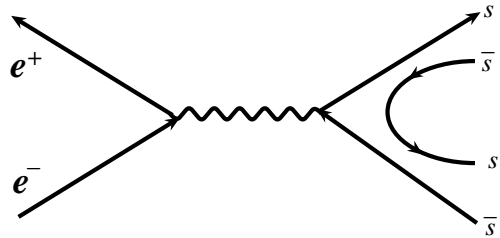}
\caption{ \small 
Possible Feynman diagram for  \epem\to$\phi\eta$.
}
\label{fig:vpdiagram}
\end{center}
\end{figure}

We report the observation of \epem\to$\phi\eta$, which is  analogous, in the  $s$ quark 
sector, to the  process    
\epem\to$J/\psi\eta_c$, since the $\eta$ meson  has an  $s \bar{s}$ quark-pair 
component. The  Feynman diagram for the most likely production mechanism 
is shown in Fig.~\ref{fig:vpdiagram}.  However, since $\eta$ is not purely  
$s \bar{s}$, the  cross section for this production mechanism is determined by  
the projection onto the   $s \bar{s}$ component of the $\eta$ meson. 
A  calculation using the QCD-based light cone method with relativistic treatment for 
the light $s$ quark is possible and therefore can provide a theoretical 
estimation~\cite{bslightcone}.

The  $\phi\eta$ combination is  a 
$\mathrm{vector-pseudoscalar} ~(\mathrm{VP})$ final state.
The production rates for \epem\to$\mathrm{VP}$ can be described by  
form factors, which are predicted  in QCD-based 
models~\cite{stanrule,Chernyak,gerard}.  
Different models predict different dependences on
    center-of-mass (CM) energy squared $s$.
The recent measurements of 
$e^+e^-\to \mathrm{VP}~(\omega\pi^0,\rho\eta~ \mathrm{and}~ \rho\eta^{\prime})$ 
from BES~\cite{bes1m0m,wang} 
investigated the $s$ dependence of the cross sections and form factors in the energy range 
from 3.65 to 3.773 GeV. 
It is interesting to investigate the $s$ dependence 
over a wider energy range.  
Since CLEO  measured the cross section 
for \epem\to$\phi\eta$ at CM energy $\sqrt{s}=3.67$ \gev~\cite{cleophieta}, a measurement 
of the same process at $\sqrt{s}=10.58$ \gev 
provides a meaningful test of the $s$ dependence.

This analysis uses  204 fb$^{-1}$  of \epem colliding beam data collected  
on the \Y4S resonance at $\sqrt{s}=10.58$ GeV and 20 fb$^{-1}$ collected 40\mev below   
the $\Upsilon(\mathrm{4S})$ mass 
with the \babar~ detector at the SLAC PEP-II asymmetric-energy $B$ factory. 
The \babar~ detector is described in detail elsewhere~\cite{babardetector}.
Charged-particle momenta and energy loss are measured in the 
tracking system that consists of   
a silicon vertex tracker (SVT) and a 
drift chamber (DCH). 
Electrons and photons  are detected in a CsI(Tl)  calorimeter (EMC). 
An internally reflecting ring-imaging Cherenkov detector (DIRC) 
provides charged particle identification (PID). 
An instrumented magnetic flux return (IFR) provides identification of muons.
Kaon and pion candidates are identified using likelihoods of   
particle hypotheses calculated from the specific ionization in the DCH and SVT 
and the Cherenkov angle measured in the DIRC. 
Photons are identified by  shower shape and lack of associated tracks.

\begin{figure}        
{\par\centering 
\resizebox*{0.48\textwidth}{!}{\includegraphics{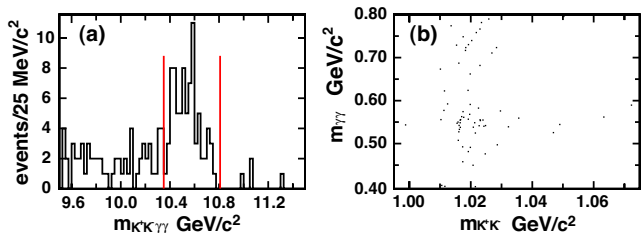} } \par}
\caption{
\label{fig:4parmass}
(a) Distribution of the invariant mass ($\Upsilon(4S)$ data) for the 
$K^+K^-\gamma\gamma$ final state near the $\phi\eta$ region.
The accepted signal region is indicated  by the lines.
(b) Scatter plot of the invariant masses of the  $K^+K^-$ and $\gamma\gamma$ pairs 
for those events in the accepted signal region.
}
\end{figure}

To reconstruct $\phi\eta$ in the $K^+K^-\gamma\gamma$ mode, 
events with exactly two well-reconstructed, oppositely charged tracks and at least  
two well-identified photons  are selected. 
Charged tracks are required to have at least 12 DCH hits and   
a laboratory polar angle within the SVT acceptance, $0.41<\theta<2.54$ 
radians. The laboratory momenta of the kaon 
candidates are required to be  greater than 
800  \mevc to reduce background. 
The  two tracks selected  must both  be identified as kaons.
We fit the two tracks to a common vertex, and require the $\chi^2$
probability to exceed 0.1\%. 
The photon candidates 
are  required to have a minimum laboratory 
energy of 500 MeV. 
The invariant mass distribution of $K^+K^-\gamma\gamma$,  after requiring the invariant mass 
of $KK$ to be near the $\phi$ mass ($m_{KK}<1.1$ \gevcc) and that of 
$\gamma\gamma$ to be near the $\eta$ mass ($0.4<m_{\gamma\gamma}<0.8$ \gevcc) is 
shown in Fig.~\ref{fig:4parmass} (a). 
We accept events with a reconstructed invariant mass of  $K^+K^-\gamma\gamma$ within 230~\mevcc of the
\epem  CM energy. There is at most one entry per event in the region of interest.

Figure~\ref{fig:4parmass}(b) shows the scatter plot of invariant masses of $K^+K^-$ and $\gamma\gamma$
pairs from the accepted \epem\to$K^+K^-\gamma\gamma$ events. 
The concentration of events indicates $\phi\eta$ production.

We use a two-dimensional log-likelihood fit to extract the signal for the reaction \epem\to$\phi\eta$. 
Due to the fact that the final state particle masses are far below
the \epem collision energy, we may treat the two-body masses as uncorrelated.  
Justified by Fig.~\ref{fig:4parmass} (b), 
the signal probability density function (PDF) is constructed as a product of two one-dimensional
PDFs, one for each resonance. 
We use a P-wave relativistic Breit-Wigner formula  to construct a PDF for the $\phi$  resonance 
and a Gaussian function to model the $\eta$ resonance.
A threshold function $q^3/(1+q^3R_t)$ is used to model the background in the  $K^+K^-$ 
system, where $q$ is the  daughter momentum in the $\phi$ 
rest frame and $R_t$ is a shape parameter. A linear function ($p_0+p_1 \cdot m_{\gamma\gamma}$) is used    
to model the background under the  $\eta$.

\begin{figure}        
{\par\centering 
\resizebox*{0.48\textwidth}{!}{\includegraphics{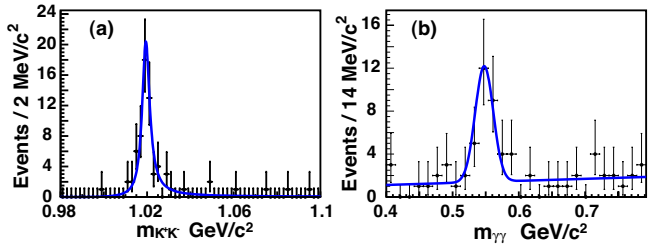}} \par}
\caption{\label{fig:massprojection}
Mass projections for (a) $K^+K^-$ pairs  and (b) $\gamma\gamma$ pairs  in   $K^+K^-\gamma\gamma$ events.
}
\end{figure}

In the fit to data, we fix the  mass and width of the $\phi$ and the mass of the 
$\eta$ to the world average values~\cite{pdg04}. The width of the $\eta$, 13.6 \mev,  is fixed to 
the resolution obtained from simulation. 
The floating parameters in the fit are: $R_t$, $p_0$, $p_1$, and the numbers of events  
for all components--$\phi\eta$, $\phi \gamma\gamma$ and $\eta K^+K^-$.  
The mass projections in $KK$ and $\gamma\gamma$ from the two-dimensional fit are shown 
in Fig.~\ref{fig:massprojection} (a)  and 
(b), respectively. We define the $\phi$ mass window  as  $1.008~<~m_{KK}~<~1.035$ \gevcc 
to reduce the systematic uncertainty due to the long tail of $\phi$ masses. 
The extracted  number of $\phi\eta$ signal events  is $24\pm 5$ in the   $\phi$  mass window,
with 20$\pm$5 in the on-resonance sample and 3$\pm$2 in the off-resonance sample. 
The number of background events within the  $\phi$ mass window and within three standard 
deviations of the $\eta$ mass   is 7$\pm$2. The significance is estimated 
by the log-likelihood difference 
between signal ($ln(L_s)$) and null ($ln(L_n)$) hypotheses (no $\phi\eta$ signal component in the PDF), $\sqrt{2 ln(L_{s}/L_{n})}$, 
which gives  6.5 standard deviations.

Given the  negative C-parity of the $\phi\eta$ final state, we assume 
$\phi\eta$ is  produced through one-virtual-photon annihilation. 
The angular distributions of  $\phi\eta$ from a $\mathrm{J}^\mathrm{P}=1^-$ initial 
state, in the helicity 
basis~\cite{chung}, can be calculated to be:
\begin{equation}
\frac{dN}{d \cos\theta^* d\cos\theta_{\phi} d\varphi_{\phi}} \propto  \sin^2\theta_{\phi}(1+\cos^2\theta^*+\cos 2 \varphi_{\phi} \sin^2\theta^*),
\label{phietaang}
\end{equation}
where the production angle $\theta^*$ is defined as the angle between the $\phi$ meson direction and 
incident $e^-$ beam in the CM frame. 
The $\phi$ helicity angle $\theta_{\phi}$  is defined as the polar angle,
measured in the $\phi$ rest frame, of the $K^+$ momentum direction with 
respect to an axis that is aligned with the $\phi$ momentum direction
in the laboratory frame.
The variable $\varphi_{\phi}$   is the  $K^+$ azimuthal angle around the direction of the $\phi$ measured with 
respect to the plane formed by  the $\phi$ and the incoming electron.
The helicity and azimuthal angles of the pseudoscalar $\eta$ are flat and thus not included 
in equation~\ref{phietaang}. Integrating over the other two angles, the distributions of 
the production 
angle, $\phi$ helicity and $\phi$ azimuthal angle are expected to be $1+\cos^2\theta^*$, 
$\sin^2\theta_{\phi}$ and $2+\cos 2 \varphi_{\phi}$, respectively. The observed angular 
distributions from \epem\to$\phi\eta$ data are consistent with the above expectation but the constraints on 
these angular distributions are  limited by statistics.

The  systematic uncertainty from the two-dimensional fit  is estimated from the difference 
in yield obtained by 
floating the mean, width, and resolution parameters in the fit.
The systematic uncertainties due to PID,  tracking, and photon 
efficiency  are estimated based on measurements
from control data samples. 
The possible 
background from related modes with an extra $\pi^0$ was estimated to be small ($<1\%$) 
by using extrapolations from statistically limited four-particle mass sidebands and we ignore it.  
The systematic uncertainties are summarized in Table ~\ref{tab:syserror}.

\begin{table}[!htb]
\caption{Systematic uncertainties on the cross section of  
$\phi\eta$.}
\begin{center}
\begin{tabular}{lrr}
\hline
Source                        & Systematic uncertainty \% \\
\hline\hline
Photon efficiency                   & 3.6     \\
Two-dimensional fit                      &  1.3     \\
Particle Identification                       &  3.0    \\
Tracking efficiency                            &  2.6     \\
 Luminosity                                     &  2.0     \\
\hline
 Total                                        &  6.0     \\
\hline\hline
\end{tabular}
\label{tab:syserror}
\end{center}
\end{table}

The radiation-corrected cross section for \epem\to$\phi\eta$ is calculated from:
\begin{equation}
\sigma=\frac{ N_{\mathrm{Observed}} }{\L  \times {\cal B}(\phi\to KK) \times {\cal B}(\eta\to \gamma\gamma)\times \varepsilon^{MC} \times (1+\delta) }
\end{equation}
where $N_{\mathrm{Observed}}$ is the extracted number of $\phi\eta$ signal events  from on- and off-resonance data, 
$\L$ is the integrated luminosity, 
${\cal B}(\phi\to KK)$ is the branching fraction of $\phi\to KK$, 
${\cal B}(\eta\to \gamma\gamma)$ is the branching fraction of $\eta\to\gamma\gamma$, 
$\varepsilon^{MC}$ is the signal efficiency obtained from Monte Carlo simulation (MC), and $\delta$ is the radiation correction 
calculated according to Ref.~\cite{radiation}. We obtain  $(1+\delta)=0.768$. The 
uncertainties due to the 
theoretical model and the $s$ dependence are negligible. The signal 
MC events are generated uniformly in phase space.  
For the determination of signal cross sections, the MC
$\cos\theta^*$, $\cos\theta_{\phi}$ and $\varphi_{\phi}$ distributions 
are re-weighted using equation~\ref{phietaang}. 
The signal efficiency in the fiducial region of $|\cos\theta^*|<$ 0.8 for
$\phi\eta$ without radiative correction is estimated to be
34.3\%, including corrections to MC
simulation for PID  and tracking. 
Taking the branching fraction of $\phi\rightarrow$$K^+K^-$ as 49.1\%, 
and $\eta\to\gamma\gamma$ as 39.4\%~\cite{pdg04},  
the final radiation-corrected cross section for  
$1.008~<~m_{\phi}~<~1.035 $ \gevcc  
within $|\cos\theta^*|<$ 0.8 near $\sqrt{s}=10.58\gev$ is: 
\begin{eqnarray*}
 \sigma_{\mathrm{fid}}(e^+e^-\rightarrow\phi \eta) & = & \Resultphieta .
\end{eqnarray*}
The cross section within $\cos\theta^* \in [-0.8,0.8]$ can be scaled to   $\cos\theta^* \in [-1.0,1.0]$
by assuming a $1+\cos^2\theta^*$ distribution to obtain: 
\begin{eqnarray*}
 \sigma(e^+e^-\rightarrow\phi \eta) & = & \Resultphietascale .
\end{eqnarray*}

To study the possibility that the observed signal is due to $\Upsilon(4S)$ 
decay, we scale the off-resonance signal to the on-resonance luminosity, and 
subtract it from the on-resonance signal. The resulting number of  
events, $-10 \pm 21$, is consistent with zero. The corresponding branching  
fraction for $\Upsilon(4S)\to \phi \eta$ is $(-0.9 \pm 1.8) \times 10^{-6}$.
Assuming this uncertainty can be treated as Gaussian and normalizing to the  
physical region ($\ge 0$),  the 90\% confidence level upper limit is $2.5  
\times 10^{-6}$.

\begin{figure}
\begin{center}
\includegraphics[width=8.5cm]{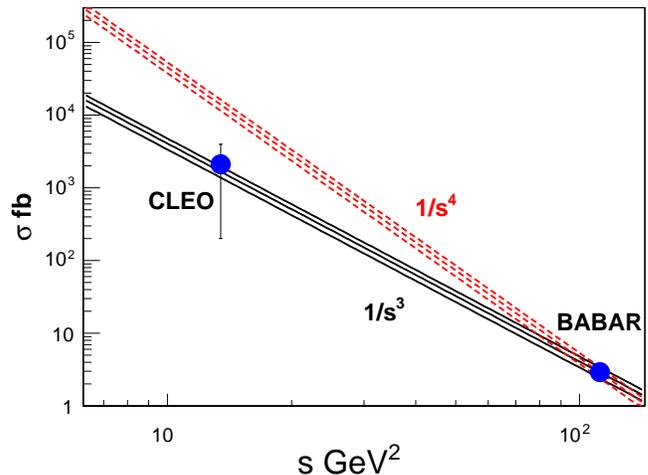}
\caption{ \small 
Cross section extrapolations 
based on \babar's measurement at $\sqrt{s}=10.58$ \gev assuming $1/s^3$ (black) or $1/s^4$ (red) 
energy dependence. The bands show one standard deviation uncertainties in the extrapolations.
The CLEO measurement at $\sqrt{s}=3.67$ \gev is also shown.
}
\label{fig:sdependence}
\end{center}
\end{figure}

There is currently no direct prediction for the cross section of this
process at this energy,
but the \epem\to$\mathrm{VP}$  cross section is expected to have $1/s^3$~\cite{gerard}
or $1/s^4$~\cite{stanrule,Chernyak} dependence in QCD-based models.
A comparison between our result and that of CLEO,
($\sigma=2.1^{+1.9}_{-1.2}\pm0.2 ~\mathrm{pb}$) at $\sqrt{s}=3.67$ GeV~\cite{cleophieta},
favors a $1/s^3$ dependence (Fig.~\ref{fig:sdependence}). We quantify the degree to which
$1/s^4$ scaling is disfavored by scaling our measured cross section in this fashion to
$\sqrt{s}=3.67$ GeV, and comparing it to the CLEO measurement. Note, however, that if
CLEO did have a downward statistical fluctuation, both their central value and their
uncertainty would be low. Accordingly, the uncertainty we use in this comparison is the CLEO
one scaled by the square root of the ratio, 2.6,  of the predicted to the observed cross sections.
The resulting disagreement with $1/s^4$ scaling is approximately 2 standard deviations.

The form of the $s$ dependence has important theoretical implications, which may affect a wide
range of QCD-based processes such as \epem\to$\mathrm{VP}$~\cite{stanrule}, 
exclusive hadronic $B$ decays~\cite{Beneke:2003zv}, and charmonium decays~\cite{Yuan:2006qz}.
The large initial-state radiation sample at the $B$ factories can
provide another route  to test the  $s$ dependence over  a wider energy range. A direct
comparison of the absolute cross section  with a possible theoretical calculation~\cite{bslightcone} 
is also interesting.

In summary, we have observed the exclusive production of $\phi\eta$ in 
$e^+e^-$ interactions at $\sqrt{s}=10.58$ GeV. 
Combining with CLEO's measurement and interpreting our result as continuum production, the 
measured $\phi\eta$ cross section favors $1/s^3$ dependence, which is in conflict with some 
QCD-based predictions. The 90\% confidence level upper limit 
on the branching fraction  $\cal{B}$$ ( \Upsilon(4S) \to \phi \eta)$ is 
  $2.5 \times 10^{-6}$.

\par We are grateful for the excellent luminosity and machine 
conditions
provided by our \pep2\ colleagues, 
and for the substantial dedicated effort from
the computing organizations that support \babar.
We wish to thank S. Brodsky, A. Goldhaber and 
G. T. Bodwin for helpful 
discussions. The collaborating institutions wish to thank 
SLAC for its support and kind hospitality. 
This work is supported by
DOE
and NSF (USA),
NSERC (Canada),
IHEP (China),
CEA and
CNRS-IN2P3
(France),
BMBF and DFG
(Germany),
INFN (Italy),
FOM (The Netherlands),
NFR (Norway),
MIST (Russia), and
PPARC (United Kingdom). 
Individuals have received support from CONACyT (Mexico), A.~P.~Sloan Foundation, 
Research Corporation,
and Alexander von Humboldt Foundation.

\end{document}